\documentclass[doublecol,linenumbers]{epl2} 
\usepackage{epsfig}
\usepackage[caption=false]{subfig}
\usepackage{graphicx}
\usepackage{dcolumn}
\usepackage{bm}
\usepackage{amsmath}
\usepackage{amssymb}
\usepackage{url}

\title{Understanding the temperature and pressure dependent electronic properties of FeSi: DFT+DMFT study}

\author{Paromita Dutta\inst{1} \and Sudhir K. Pandey\inst{2}}
\shortauthor{P. Dutta \etal}

\institute{                    
  \inst{1} School of Basic Sciences, Indian Institute of Technology Mandi, Kamand, Himachal Pradesh-175075, India\\
  \inst{2} School of Engineering, Indian Institute of Technology Mandi, Kamand, Himachal Pradesh-175075, India.
}

\abstract{
Electronic structures of FeSi and Fe$_{1.02}$Si$_{0.98}$ under pressure (achieved through volume compression) have been investigated by using DFT+DMFT and KKR-CPA methods, respectively. The widening of band-gap with increasing pressure suggests that the experimentally observed insulator to metal transition temperature should shift towards the higher temperature for FeSi. KKR-CPA calculations have shown the presence of impurity states in the gapped region which predicts the half-metallic nature. The closure of gap (in one spin channel) with pressure increment appears to be responsible for experimentally observed semiconductor to metal transition in Fe excess samples at a temperature below 50 K. Magnetic moments at Fe excess sites are found to be decreasing with increasing pressure from 2.4 $\mu_B$ per Fe atom (612 Bohr$^3$) to 1.2 $\mu_B$ per Fe atom (507 Bohr$^3$). Moreover, for FeSi the calculated local spin susceptibility has shown decreasing behavior with pressure rise similar to experimental result. 
}

\begin{document}

\maketitle

\section{Introduction}
\small
FeSi belongs to the family of transition metal silicides which crystallizes into noncentrosymmetric cubic B20 structure. This family possesses variety of complex phenomena which have applications in the field of applied sciences \cite{Manyala,Keller,Schulz}. Amongst these FeSi has the potential for the applications in electronics, spintronics and thermoelectrics \cite{Lange,Slack}. Likewise, Sales \textit{et al.} \cite{Sales} have shown its applicability as a thermoelectric material in refrigeration. It has been under study for decades due to its temperature dependent magnetic and electrical properties \cite{V,Wertheim,Watanabe,Moriya,Petrova,Sakai,Imada,Shu,Wolfe}. Since its discovery FeSi has been a controversial material due to its unusual physical properties. One of its peculiar property is its magnetic susceptibility ($\chi(T)$), which increases with temperature rise after 100 K and has a broad maximum around $\sim$ 500 K and then it drops; follows Curie-Weiss law at higher temperature \cite{V}. However, absence of spin ordering at lower temperatures is observed as its $\chi(T)$ vanishes below 50 K \cite{Watanabe}. Moreover, its $\chi(T)$ value decreases rapidly as the temperature is lowered below 500 K and thought it to be the onset of anti-ferromagnetic behavior below this temperature \cite{V}. However, neutron scattering, M\"{o}ssbauer, and nuclear magnetic resonance measurements showed no long-range magnetic order \cite{Watanabe,Wertheim}. 

Another peculiar property is its electrical conductivity ($\sigma$) whose measurement has shown different temperature regimes of transport while exhibiting an insulating ground state. Firstly, the semiconductor to metal transition at $\sim$ 200 K \cite{Chernikov,Aarts,Petrova,Sakai,Shu}. Secondly, the transport behavior at low temperature range ($\sim$ 5-40 K) is of variable hopping range (VRH) type and has shown the existence of localized states in the gapped region \cite{Takagi,Lisunov,Mani2004,Hearne,Mani,Chernikov}. Besides these studies, the temperature-dependent properties are still elusive and consequently the controversies are continued \cite{Tomczak,Tomczak2018}. Thus, considering the ambiguities associated with the temperature-dependent properties of FeSi many electronic structure calculations and measurements are carried out \cite{Mattheiss,Tomczak,Aeppli,Mazurenko,Dutta,Doniach,J,Klein}. Out of these, density functional theory + dynamical mean field theory (DFT+DMFT), advance studies have shown the essential effects of hybridization on Fe - 3\textit{d}/Si -3\textit{p} states in FeSi at DFT level while crucial effect of electronic correlations at DMFT level (where temperature rise results in a transition from a nonmagnetic insulator to a bad metal) \cite{Tomczak,Aeppli,Mazurenko}. Likewise, a pseudogapped region with strongly renormalized density of states (DOS) at the Fermi level (E$_F$) consisting of upper and lower Hubbard bands with $\pm U/2$ \cite{Mazurenko}; showing the effect of electronic correlations on the system. Here, it is important to note that in ref\cite{Mazurenko}, the  presence of lower and upper Hubbard bands do not predict FeSi as Mott insulator. Instead their results have clearly shown that FeSi is a correlated band insulator. Similarly, our earlier work has also shown the importance of electronic correlations in understanding the physical properties of FeSi \cite{Dutta}. Experimental measurements have also shown the existence of the narrow peaked DOS around E$_F$ \cite{Doniach,J,Klein}. All these studies suggest that both hybridization and electronic correlations are playing an important role in deciding the unusual physical properties of FeSi system.

FeSi being one of the members of strongly correlated electron systems. In this family of materials physical properties are extremely sensitive to small changes in external parameters such as temperature, pressure, electronic composition, electric/magnetic field. In consequence of which their sensitivity towards external parameters is very useful in exploring the materials of this class. In view of this, many doping studies have been made over FeSi system \cite{Beille,Chernikov,Manyala,Bauer1998,Bauer,Sales,Sales2011}. Likewise, pressure has also been taken as a probe to account the effect of hybridization and electronic correlations on the transport behavior of FeSi and doped FeSi systems \cite{Bauer,Reichl,Pelzer,Mani,Hearne,Koyama,Grechnev}. After FeSi is studied under pressure, two temperature regimes are again found where contradictory results of $\rho$ and E$_g$ are showing up. Following Reichl \textit{et al.} work \cite{Reichl} where below 50 K the electrical resistivity ($\rho$) is decreasing with pressure (applied upto 9.3 GPa) while in the range 100 - 300 K $\rho$ is increasing. Moreover, Bauer \textit{et al.} \cite{Bauer} and Reichl \textit{et al.} \cite{Reichl} have shown the widening of the gap width with pressure. However, following the work of Pelzer \textit{et al.} \cite{Pelzer}, here the E$_g$ is increasing with pressure upto $\sim$ 1 GPa but after that it keeps decreasing until reaches 8 GPa for Fe$_{1.003}$Si$_{0.997}$ material. Interestingly, Mani \textit{et al.} \cite{Mani} have shown the decreasing and increasing trends of the E$_g$ for nearly stoichiometric FeSi samples ( one is Fe excess where Fe = 50.02 at.\% and other is Fe deficient material where Fe = 49.9 at.\%) when pressure is upto 6.4 GPa and 3.5 GPa, respectively. Recently, Hearne \textit{et al.} \cite{Hearne} have also reported the decreasing behavior of E$_g$ with pressure upto 32.1 GPa for samples of nearly stoichiometry i.e., Fe$_{1.01}$S$_{0.99}$. Moreover, at temperature $<$ 50 K the decreasing behavior of $\rho$ when pressure is applied upto 11.2 GPa while semiconductor to metal transition at pressure (P) $\geq$ 15 GPa are also observed.

All these experimental studies are showing the disparate behavior of $\rho$ and E$_g$ under pressure at low and high temperatures corresponding to off-stoichiometry of FeSi. Generally, the samples prepared experimentally are often off-stoichiometric and may be the reason for seeing such odd results. In light of this, we have carried out electronic structure calculations for both FeSi and Fe$_{1.02}$Si$_{0.98}$, respectively, with pressure aspect. This pressure aspect is considered by reducing the lattice parameter of experimental structure. We have used the DFT+DMFT advanced method \cite{Yee} for the analysis of FeSi. Our DFT+DMFT calculations have revealed the increasing trend of E$_g$ with increase in pressure which is similar to experimental observation for stoichiometric FeSi. We have used the DFT based Korringa-Kohn-Rostoker (KKR) under the coherent potential approximation (CPA) \cite{kkr} for the analysis of Fe$_{1.02}$Si$_{0.98}$. From KKR-CPA we found that the impurity states have been generated in the gapped region around the E$_F$ and they have half-metallic behavior. With volume decrement the closure of energy gap (in one channel) appears to be responsible for the experimental observation of semiconductor to metal transition at temperature below 50 K for Fe excess sample. Magnetic moments at Fe excess sites are calculated, and they are found to be decreasing with volume compression. For FeSi, the calculated local spin susceptibility is decreasing with volume compression as observed experimentally. Moreover, from the imaginary part of impurity hybridization function (- Im $\Delta$), the degree of localization seems to decrease with volume compression in FeSi. Contradictorily temperature rise appears to localize Fe 3\textit{d} electrons in FeSi. 

\small
\section{Computational Details}

The electronic structure calculations with the inclusion of spin-orbit coupling (SOC) for stoichiometric FeSi by using DFT and DFT+DMFT methods have been carried out. Firstly, DFT calculations are performed by using the \textit{state-of-the-art} full-potential linearized augmented plane-wave (FP-LAPW) method accomplished by WIEN2k code \cite{Blaha}. The initial lattice constants are taken from the literature \cite{Ark} for the calculations. The exchange-correlation functional local density approximation (LDA) is chosen here \cite{Jones}. The muffin-tin sphere radii of 2.18 Bohr and 1.84 Bohr for Fe and Si sites, respectively, are used. 1000 $k$-points mesh grid size in whole Brillouin zone has been used. In this part of the calculations, volume-optimization of the structure is also carried out. For this, the calculated lattice constants (as provided in Table I) are computed by fitting total energy versus unit cell volume data with the Birch-Murganhan (BM) equation of state (eos) \cite{Birch}. Third-order BM isothermal eos is given in the Eq. 1; pressure is also computed for every volume from the Eq. 2 \cite{Birch}:

\begin{equation}
\begin{split}
 {E(V) = E_0 + \frac{9V_0B_0}{16} \Big[\Big\lbrace\Big(\frac{V_0}{V}\Big)^{2/3}-1\Big\rbrace^3 B'_0} \\{ + \Big\lbrace\Big(\frac{V_0}{V}\Big)^{2/3}-1\Big\rbrace^2 \Big\lbrace 6 - 4\Big(\frac{V_0}{V}\Big)^{2/3} \Big\rbrace \Big]}
\end{split}
\end{equation}

\begin{equation}
\begin {split}
{P(V) = \frac{3}{2} B \Big\lbrace\Big(\frac{V_0}{V}\Big)^{7/3} - \Big(\frac{V_0}{V}\Big)^{5/3}\Big\rbrace \Big[1+ } \\ { \frac{3}{4} \Big(B'_0 - 4\Big) \Big\lbrace\Big(\frac{V_0}{V}\Big)^{2/3} - 1 \Big\rbrace\Big]}
\end{split}
\end{equation}
where, E is energy, V is volume, $B_0$ is equilibrium bulk modulus, $V_0$ is volume of unit cell corresponding to minimum energy, $B'_0$ is pressure derivative of bulk modulus at equilibrium value and P is pressure. The process of volume reduction is carried out by varying lattice constants. Next, spin-polarized calculation of onsite Coulomb interaction (\textit{U$_{eff}$}) for different unit cell volumes by using constrained DFT method \cite{Anisimov,Madsen,Blaha} is carried out. We have followed the same procedure for evaluating \textit{U$_{eff}$} for different cell volumes as given in our earlier works \cite{Paromita,Lal,Pandey,S}. Here, it is important to note that the evaluated values of \textit{U$_{eff}$} for different cell volumes are almost same as previously \cite{Paromita} calculated for the unit cell volume of 612 Bohr$^3$. A difference of $\sim$ 0.2 eV is found, and thus the self-consistently calculated values of \textit{U$_{eff}$} \& \textit{J} with 4.4 eV \& 0.89 eV, respectively, are used \cite{Paromita}.

Secondly, DFT+DMFT calculations are carried out by the usage of WIEN2k code \cite{Blaha} and the code as implemented by Haule \textit{et al.} \cite{Yee}. Here, DFT+DMFT functional is implemented in the real space embedded DMFT approach \cite{Yee}, which delivers stationary free energies at finite temperatures \cite{Birol}. 
DMFT calculations are performed for 100 K and 300 K temperatures, and these calculations are fully self-consistent in electronic charge density and impurity levels. For solving the auxiliary impurity problem, a continuous-time quantum Monte Carlo impurity solver has been used \cite{K}. The scheme for exact double-counting as proposed by Haule has been used here \cite{Haule}. More informations regarding this DFT+DMFT code as developed by Haule \textit{et al.}\cite{Yee} can be found at references\cite{Yee,Paro,Kotliar,dmft}. Fe 3\textit{d} orbitals are treated at DMFT level. For the calculations, the density-density form of the Coulomb repulsion has been used. All these DMFT calculations are converged on the imaginary axis. Then, for obtaining the self-energy on the real axis, an analytical continuation is needed to be done. This analytical continuation is achieved by using maximum entropy method \cite{Jarrell} for the spectra on the real axis. 2000 $k$-points grid is used for the density of states (DOS) calculations on both DFT and DMFT levels. 

Lastly, magnetic calculations for Fe$_{1.02}$Si$_{0.98}$ by using Korringa-Kohn-Rostoker (KKR) under the coherent potential approximation (CPA) in the framework of LDA are carried out \cite{kkr}. 76 number of \textit{k}-points are used in the irreducible part of the Brillouin zone. Same muffin-tin radii of Fe and Si are used here.

\section{Results and Discussion}

\begin{figure}[tbh]
  \begin{center}
    \includegraphics[width=2in]{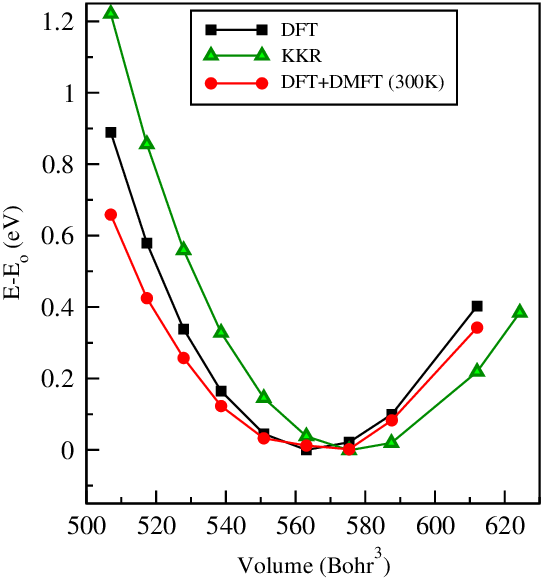}
  \end{center}
  \caption{\small{(color online) Energy versus volume curve for pure FeSi with spin-orbit coupling for three energies as obtained from DFT, KKR and DFT+DMFT$_{300K}$ calculations. }}
 \end{figure}

\begin{table}[tbh]
\caption{\footnotesize{Calculated lattice constants and Bulk modulus of FeSi for different evaluated energies.}}
\label{tab.1}
\begin{center}
\setlength{\tabcolsep}{5pt}
\footnotesize
\begin{tabular}{lccccr}
\hline
\hline
Method  & lattice constant (a$_{0}$)  & Bulk modulus (B$_{0}$) \\
  & (\text{\AA}) & (GPa)\\
\hline 
DFT & 4.3736 & 260.58 \\
KKR & 4.4035 & 246.02\\
DFT+DMFT$_{300K}$ & 4.3781 & 223.61 \\
\hline
\hline
\end{tabular}
\end{center}
\end{table}

In Fig. 1, the obtained values for total energy difference of volume dependent energies and equilibrium volume energy (E-E$_{0}$) are plotted as a function of volume for stoichiometric FeSi only. In this figure, three total energies as plotted for FeSi are calculated by using DFT, KKR and DFT+DMFT (at T = 300 K) methods, respectively. The calculated data are then fitted by using BM eos \cite{Birch}. The calculated values of lattice constant (a$_{0}$) and Bulk modulus (B$_{0}$) corresponding to energies evaluated from different methods are tabulated in Table I. From the table, we can see that all the methods are giving almost similar a$_{0}$ value of $\sim$ 4.4 \text{\AA} whereas experimental a$_{0}$ values at 300 K are found to vary in the range of $\sim$ 4.46 to 4.48 \text{\AA} \cite{Ark,knittle,lin,wood}. For the B$_{0}$ values in the Table I, it is observed that DFT+DMFT$_{300K}$ has estimated its value of $\sim$ 223.6 GPa which is the closest to the experimentally found B$_{0}$ values ranging from $\sim$ 115-209 GPa at 300 K \cite{knittle,lin,wood,guyot,Zino} out of all the methods used here. Thus, DFT+DMFT happens to provide quite appropriate value for B$_{0}$. Next, in Table II we have tabulated the calculated values of pressure as computed from BM eos \cite{Birch} for some of the volumes for which the study has been done. We have presented the study with respect to volume due to the difference found in the calculated values of pressure corresponding to different methods for a particular unit cell volume. However, pressures corresponding to DFT+DMFT method can be helpful to the experimentalist if tries to perform any pressure induced experiment due to the fair closeness of B$_{0}$ with its experimental results.

\begin{table}
\caption{\footnotesize{Calculated pressures by using BM equation state \cite{Birch} of FeSi for the reduced unit cell volumes from its experimental volume = 612 Bohr$^{3}$.}}
\label{tab.2}
\begin{center}
\setlength{\tabcolsep}{0.5pt}
\footnotesize
\begin{tabular}{lcccr}
\hline
\hline
  &  & & Pressure &  \\
 \cline{3-5}
\\
  Unit cell volume &  lattice constant &  DFT  &  KKR & DFT+DMFT$_{300K}$ \\
(Bohr$^{3}$) & (Bohr) & (GPa) & (GPa) & (GPa)\\
\hline 
 
507 & 7.97 &36.73 & 42.70 & 27.04   \\
517 & 8.03 &28.47 & 34.38 & 21.90\\
528 & 8.08 &20.94  & 26.75 & 16.85  \\
539 & 8.14 &14.04  & 19.72& 11.91\\
551 & 8.20 &7.08 & 12.58& 6.56\\
563 & 8.26 &0.94  & 6.25 & 1.51\\
575 & 8.32 &-4.48 & 0.63 & -3.27 \\
588 & 8.37 &-9.22 & -4.33 & -7.70\\
612 & 8.49 & -17.18 & -12.75 & -15.98 \\
\hline
\hline
\end{tabular}
\end{center}
\end{table}

\begin{figure}
  \begin{center}
    \includegraphics[width=2.2in]{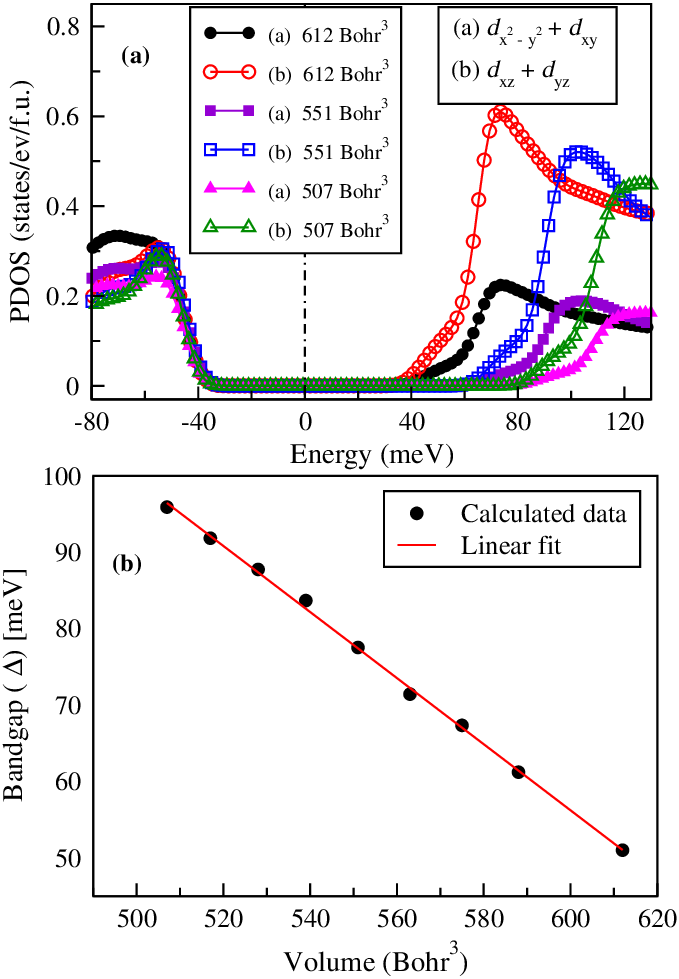}
  \end{center}

  \caption{\small{(color online) (a) Partial density of states (PDOS) of the orbitals \textit{d$_{x^{2}-y^{2}}$} + \textit{d$_{xy}$} and \textit{d$_{xz}$} + \textit{d$_{yz}$} of Fe-3\textit{d} states calculated within DFT for three different unit cell volumes 612, 551 and 507 Bohr$^{3}$; zero energy corresponds to the Fermi level (dotted line), and (b) Band gap versus different unit cell volumes for the range 612-507 Bohr$^{3}$ of FeSi.}}
\end{figure}

Next, we would like to see the effect of reduced unit cell volumes over the E$_g$ of FeSi when studied at DFT level. Accordingly, we have shown a plot of partial density of states of the orbitals \textit{d$_{x^{2}-y^{2}}$} + \textit{d$_{xy}$} and \textit{d$_{xz}$} + \textit{d$_{yz}$} of Fe-3\textit{d} states calculated within DFT for three different unit cell volumes = 612, 551 and 507 Bohr$^{3}$ in Fig. 2(a). 612 Bohr$^{3}$ is the volume of experimentally realized structure of FeSi \cite{Ark}. From the plot, it is observed that with volume decrement the orbitals \textit{d$_{x^{2}-y^{2}}$} + \textit{d$_{xy}$} and  \textit{d$_{xz}$} + \textit{d$_{yz}$} are moving apart from the E$_F$, and increasing their corresponding energy gaps. For example, the gap for the vol = 612 Bohr$^3$ is $\sim$ 51 meV has increased to $\sim$ 95.9 meV when the vol = 507 bohr$^3$. This suggests the existence of large overlapping between these orbitals which further results in increasing the gap between them. This can be related with process of hybridization where large overlapping results in increasing the gap between the bonding and anti-bonding orbitals. Thus, suggesting that the material is a simple semiconductor at 0 K where hybridization seems to be responsible for the existence of E$_g$. We have also calculated the band-gaps corresponding to unit cell volumes ranging from 612 to 507 Bohr$^{3}$ and plotted them in Fig. 2(b). A linear fit of the calculated data corresponding to band-gaps has been done. From the linear fit the rate of decrement of the E$_g$ is of $\sim$ 0.43 meV/Bohr$^3$; showing the widening of the narrow E$_g$ with volume compression. Further, by extrapolating this data we have also calculated a specific unit cell volume of $\sim$ 730 Bohr$^{3}$ which corresponds to zero band-gap. Zero E$_g$ can be interpreted as a signature of weak hybridization due to increase in the volume to $\sim$ 730 Bohr$^3$.

\begin{figure}
  \begin{center}
    \includegraphics[width=2.3in]{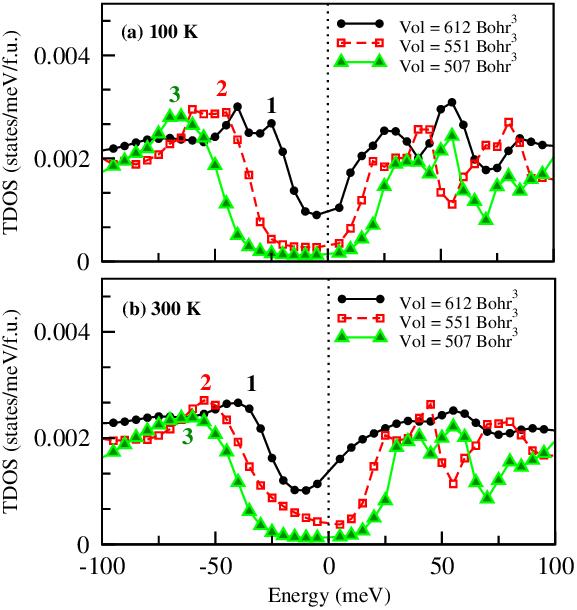}
  \end{center}

  \caption{\small{(color online) Total density of states (TDOS) of FeSi calculated within DFT+DMFT for three different unit cell volumes 612, 551 and 507 Bohr$^{3}$ for temperatures \textit{viz.} (a) 100 K, and (b) 300 K, respectively. Zero energy corresponds to the Fermi level (dotted line).}}
\end{figure}

Now in this part of discussion, we have shown the changes occurred in the interacting total density of states (TDOS) of FeSi as calculated within DFT+DMFT with reduction in unit cell volumes (or increase in pressure) and temperature simultaneously. 
Fig 3 has been drawn showing the calculated interacting TDOS of FeSi for three different unit cell volumes 612, 551 and 507 Bohr$^{3}$ for temperatures \textit{viz.} (a) 100 K and (b) 300 K, respectively. For the reference, we have marked few peaks as 1, 2, and 3 lying in valence band (VB), where they correspond to vol = 612, 551 and 507 Bohr$^{3}$, respectively. The changes we have found in interacting TDOS with volume decrement while going from Fig. 3(a) to 3(b) are (i) DOS around the E$_F$ is diminishing at each temperature and (ii) flattening of the energy peaks in conduction band (CB) further changing the energy positions of peaks 2 and 3 in VB corresponding to 551 and 507 Bohr$^3$ at 300 K.

It is well known the understanding of DOS provides a good insight of the expected transport behavior of any material. Knowing this we would like to understand the experimental electrical resistivity ($\rho$) data  of FeSi (measured with/without application of pressure) \cite{Petrova,Sakai,Shu,Spolenak,Bauer,Reichl,Pelzer,Mani,Hearne} with the help of TDOS plot in the fig 3. Before this, we need to understand a simple expression of conductivity given in Eq. (3) and its relative change ($ \Delta \sigma$) with respect to temperature is given in Eq. (4).

\begin{equation} 
\sigma = \frac{ne^2\tau}{m}
\end{equation}

\begin{equation}
\frac{\Delta \sigma}{\sigma}= \frac{\Delta n}{n} + \frac{\Delta \tau}{\tau}
\end{equation}

where, $e$ is electronic charge, $m$ is mass of the charge carrier, $n$ is the charge carrier concentration, $\tau$ is relaxation time, $\Delta n$ and $\Delta \tau$ are the changes in the charge carrier concentration and relaxation time, respectively, with respect to temperature. As states in the range of $k_BT$ around the E$_F$ participate in the transport of charge carriers. Thus in Eq. (3), with rise in temperature $n$ will always increase while $\tau$ will always decrease. Consequently, $\Delta \tau$ will always be negative with temperature increase while $\Delta n$ is expected to be positive with temperature rise. Thus, when $|\frac{\Delta n}{n}|>|\frac{\Delta \tau}{\tau}|$ with temperature rise then $\frac{\Delta \sigma}{\sigma} > 0$, and the material will behave as semiconductor. On other hand, when $|\frac{\Delta n}{n}|<|\frac{\Delta \tau}{\tau}|$ with temperature rise, then $\frac{\Delta \sigma}{\sigma} < 0$ thereby making the material to behave as metal. However, if $|\frac{\Delta n}{n}| >>|\frac{\Delta \tau}{\tau}|$ with temperature rise, the $\frac{\Delta \sigma}{\sigma}$ is expected to be positively large. The similar changes in the experimental $\rho$ is seen from $\sim$ 2.5 to 200 K \cite{Petrova,Sakai,Shu,Spolenak} and suggesting FeSi to be semi-conducting. Thus, in Fig. 3(a) the presence of sharp edges around the E$_F$ appears to be responsible for making $|\frac{\Delta n}{n}|>>|\frac{\Delta \tau}{\tau}|$ with per Kelvin rise in temperature around 100 K which is for 612 Bohr$^3$. However, for 551 Bohr$^3$ at 100 K the deep well has further gone down with lesser interacting DOS if compared to 612 Bohr$^3$. As a result FeSi at 551 Bohr$^3$ is expected to remain in its insulating state. Similar insulating behavior of FeSi under pressure (upto $\sim$ 10 GPa) around 100 K has also been witnessed by experimental $\rho$ data \cite{Bauer,Reichl} which is in our case it is upto 539 Bohr$^3$ ($\sim$ 12 GPa according to Table II). Further moving to 507 Bohr$^3$, the deepening of well continues with negligible interacting DOS, suggesting an insulating behavior to be followed leading to the formation of band-gap $\sim$ 52 meV at 100 K with further volume decrement.

Now on looking at the Fig. 3(b), shallowing of DOS near to E$_F$ has emerged for 612 Bohr$^3$. This flattening of interacting DOS will tend to reduce the value of $\frac{\Delta n}{n}$ with per Kelvin rise in temperature around 300 K, resulting in lowering the difference between $\frac{\Delta n}{n}$ and $\frac{\Delta \tau}{\tau}$ with increase in temperature. As a result it is now expected to follow the relation of $|\frac{\Delta n}{n}|>|\frac{\Delta \tau}{\tau}|$, and this will make $\frac{\Delta \sigma}{\sigma}$ to decrease around 300 K. The similar behavior is also observed in the experimental $\rho$ data \cite{Petrova,Sakai,Shu,Spolenak}, and thus suggesting the material to be less insulating than 100 K. However, at 551 Bohr$^3$ the deepening of edge has started again with almost negligible interacting DOS. Thus, for 551 Bohr$^3$ around 300 K the material is expected to be in its semi-conducting state. Furthermore for 507 Bohr$^3$, the deepening of edges continues with volume decrement while the interacting DOS remains negligible; suggesting it to remain in its semi-conducting state with band-gap $\sim$ 55 meV. Thus, with volume decrement the material becomes more insulating. This seems to be in accordance with experimental works of Bauer \textit{et al.}\cite{Bauer}, Reichl \textit{et al.} \cite{Reichl} and Koyama \textit{et al.} \cite{Koyama}, where they have reported the increment in band-gap when pressure is increased upto 1.3 GPa, 9.3 GPa and 1.2 GPa, respectively. Similar results also shown by Mani \textit{et al.} \cite{Mani} where the band-gap has increased with pressure upto 3.5 GPa for FeSi (with poor resistivity ratio (RR = 7) where Fe = 49.9 at. \%). Hence, it appears that with decreasing volume the experimentally observed semiconductor to metal transition temperature is expected to shift towards higher temperature. This kind of result where widening of gap under pressure at higher temperatures is also reported in the work of Cooley \textit{et al.} \cite{Cooley}.

\begin{figure}[tbh]
  \begin{center}
    \includegraphics[width=3.2in, height=2in]{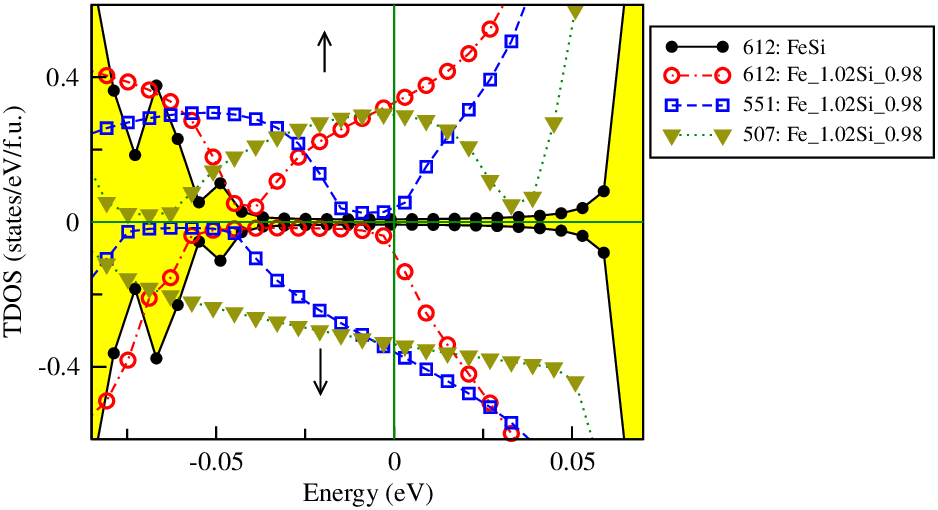}
  \end{center}

  \caption{\small{(color online) Total density of states of FeSi (shaded region) and Fe$_{1.02}$Si$_{0.98}$ calculated by using KKR - CPA \cite{kkr} for 612 and 612-551-507 Bohr$^3$ unit cell volumes for both up and down spins, respectively.}}
\end{figure}

Based on our above discussions, it can be said that for FeSi conductivity will decrease with volume compression. However, recently in Hearne \textit{et al.} work \cite{Hearne} at T $<$ 50 K, resistivity data have shown decreasing behavior upto 11.2 GPa and when pressure $\geq$ 14.4 GPa  an insulator to metal transition is witnessed for nearly stoichiometric FeSi sample (which is Fe$_{1.01}$Si$_{0.99}$). Similarly, Mani \textit{et al.} \cite{Mani} have reported for Fe excess sample (Fe = 50.02 at.\%); composition close to stoichiometry. Their resistivity curve also shows decreasing behavior at lower temperatures when pressure is applied upto 6.4 GPa \cite{Mani}. These two works have reported similar resistivity behavior for Fe excess samples upto pressure 6.4 GPa. In order to understand the experimental resistivity behavior we have plotted a TDOS for pure FeSi (stoichiometric) and Fe$_{1.02}$Si$_{0.98}$ (off-stoichiometric) computed by using DFT based KKR \cite{kkr} and KKR-CPA methods \cite{kkr}, respectively, for up and down spins as shown in Fig 4. In this figure, for pure FeSi corresponding to 612 Bohr$^3$ a band-gap of $\sim$ 76 meV is found. On the other hand, for Fe$_{1.02}$Si$_{0.98}$ with the vol = 612 Bohr$^3$ impurity states are generated in the gapped region around the E$_F$. These states are mainly contributing from the up channel while a small energy gap of $\sim$ 35 meV is found in the down channel. The presence of these impurity states around the E$_F$ may give rise to its metallic behavior but experimentally an insulating behavior is observed \cite{Hearne,Mani}. The impurity states are normally localized which may be the reason here that they are not reflecting in the transport measurement. Similarly, experimental works have recognized those impurity states as localized states in the gapped region \cite{Takagi,Lisunov}. 

Further moving to 551 Bohr$^3$, we found that the gap has now reduced to $\sim$ 18 meV in the up channel while maximum states contribute from down channel. Due to a substantial decrement in the energy gap, the resistivity should decrease here. Similar decreasing behavior of resistivity is reflected upto 11.2 GPa in the experiment \cite{Hearne}. On comparing this range of experimental pressure with the calculated pressures (as tabulated in Table II) corresponding to 551 Bohr$^3$, we can see DMFT result ($\sim$ 6.56 GPa) is the most closest one here \cite{Hearne,Mani}. Next moving to 507 Bohr$^3$, the energy gap has completely disappeared and the impurity states are now contributing from both up and down channels collectively around the E$_F$. It seems that at this volume the material is expected to be metallic. Moreover, at pressure $\geq$ 14.4 GPa a metallic phase has also been observed experimentally \cite{Hearne}. This suggests that between 551 to 507 Bohr$^3$ there is a possibility of observing a closure of the energy gap and an insulator to metal transition to occur here. The pressure calculated from DMFT corresponding to 507 Bohr$^3$ $\sim$ 27.04 GPa is much higher than 14.4 GPa; indicating that at 507 Bohr$^3$ the presence of immense impurity states around the E$_F$ will account to its metallic behavior. Hence, it appears that with volume compression the closure of the energy gap is leading to experimental observation of semiconductor to metal transition at T $<$ 50 K \cite{Hearne}. Similar DOS behavior with volume reduction is also witnessed when KKR-CPA calculations are performed for Fe$_{1.01}$Si$_{0.99}$ composition. Hence, it can be said that off-stoichiometry in FeSi is prerequisite for observing metalization at high pressure. Here, it is important to note that Fang \textit{et al.} have experimentally found the conducting surfaces in FeSi when T $<$ 19 K\cite{Fang}.

KKR calculation for FeSi has also shown its non-magnetic ground state (as expected) whereas KKR-CPA calculations have shown magnetic ground state for Fe excess sample. The calculated magnetic moments at Fe excess sites for Fe$_{1.02}$Si$_{0.98}$ at the volumes 612, 551 and 507 Bohr$^3$ are found to be $\sim$ 2.4, $\sim$ 1.8 and $\sim$ 1.2 $\mu_B$ per Fe atom, respectively. These magnetic moments are decreasing with volume compression. For better understanding we have also tried to find the critical value of deviation from stoichiometry after which magnetism may seem to emerge. For this magnetic KKR-CPA calculations are carried out for Fe$_{1.0001}$Si$_{0.9999}$ composition; resulted in net magnetic moment. This indicates that such a small amount of excess Fe is capable of creating a magnetic moment at Si site. Yeo \textit{et al.} have experimentally found a ferromagnetic state (FM) in FeSi$_{1-x}$Ge$_x$ when x $\geq$ 0.25 \cite{Yeo}. To verify whether the existence of magnetic state is the intrinsic property of the material or the result of off-stoichiometry, we carried out KKR-CPA calculations for FeSi$_{1-x}$Ge$_x$ (0.25$\leq$x$\leq$1). The calculations have shown their non-magnetic ground state. This suggests that the experimental observation of FM ground state in the material is as a result of excess Fe at Si site which may be  below detection level. Hence, all these results suggest that strong Hund’s coupling must be responsible for creation of net magnetic moment at Si site. 

\begin{figure}[tbh]
  \begin{center}
    \includegraphics[width=2in, height=2in]{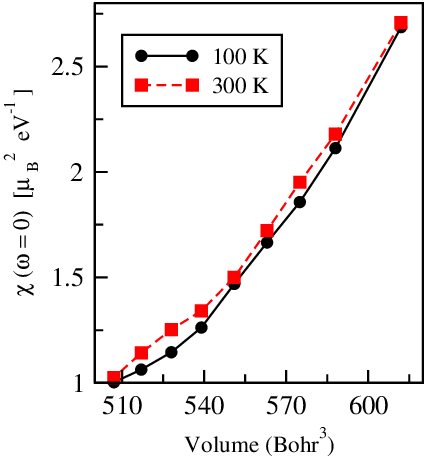}
  \end{center}

  \caption{\small{(color online) Local spin susceptibility ($\chi$) at $\omega$ = 0 (calculated within DFT+DMFT) versus unit cell volumes ranging from 612 to 507 Bohr$^{3}$ for 100 K and 300 K, respectively.}}
\end{figure}

In the last part of results and discussion, we have presented the plot for the calculated local spin susceptibility ($\chi$) at $\omega$ = 0 versus unit cell volumes ranging from 612 to 507 Bohr$^{3}$ for 100 K and 300 K, respectively, in Fig. 5. This $\chi$ is basically providing the information of the local spin-spin correlations. From the plot we have found that $\chi$ for 612 Bohr$^{3}$ volume has a value of $\sim$ 2.68 $\mu_B^2 eV^{-1}$ at 100 K and $\sim$ 2.7 $\mu_B^2 eV^{-1}$ at 300 K. The figure shows a decreasing trend of $\chi$  value with reducing unit cell volumes. The similar decreasing trend of magnetic susceptibility with increasing pressure is seen in the experimental work \cite{Koyama}. Moreover, it is well known that DMFT is capable of providing the informations regarding quasiparticles’ lifetime (from self-energy calculation) and interactions of the correlated electrons with the bath (from impurity hybridization function). Thus, these insightful informations are given in Sec S1. and Sec S2. of the supplementary material \cite{SP}.

\small
\section{Conclusion}

In this work, we have presented the effect of pressure (attained through volume decrement) on the electronic structure of FeSi and Fe$_{1.02}$Si$_{0.98}$ by using DFT+DMFT and KKR-CPA methods, respectively. At 300 K, we have calculated values of lattice constant $\sim$ 4.4 \text{\AA} and Bulk modulus $\sim$ 223.6 GPa, which are closely matched with the experimental results. DFT+DMFT calculations have revealed the band-gap increment with volume reduction which is in accordance with experimental observations of stoichiometric FeSi. This suggests that the insulator to metal transition temperature should shift towards higher temperature with volume compression. KKR-CPA calculations have shown the presence of impurity states in the gapped region around the E$_F$ which is half-metallic in nature. The closure of energy gap (in one spin channel) with volume decrement seems to be responsible for the experimental observation of insulator-metal transition at temperature below 50 K for Fe excess samples. 
The calculated magnetic moments at Fe excess sites for Fe$_{1.02}$Si$_{0.98}$ are found to be decreasing with volume reduction. For instance, the magnetic moments for the volumes 612, 551 and 507 Bohr$^3$ are calculated as $\sim$ 2.4, 1.8 and 1.2 $\mu_B$ per Fe atom, respectively. Moreover, for FeSi a decreasing trend in the calculated local spin susceptibility with volume compression is similar to experimental result. At the end, -Im $\Delta$ has shown that with volume reduction the degree of localization of Fe 3\textit{d} electrons is decreasing in FeSi. However, -Im $\Delta_{300K} <$ -Im $\Delta_{100K}$ (magnitude wise) for all the reduced volumes are observed; suggesting the localization of Fe 3\textit{d} electrons with temperature rise.










\bibliography{manuscript_epl}

\begin{thebibliography}{10}
\expandafter\ifx\csname url\endcsname\relax\def\url#1{\texttt{#1}}\fi

\bibitem{Manyala}
\Name{{Manyala} N., {Sidis} Y., {DiTusa} J.~F., {Aeppli} G., {Young} D.~P. \and
  {Fisk} Z.} \REVIEW{Nature}{404}{2000}{581}.

\bibitem{Keller}
\Name{Pfleiderer C., B{\"o}ni P., Keller T., R{\"o}{\ss}ler U.~K. \and Rosch
  A.} \REVIEW{Science}{316}{2007}{1871}.

\bibitem{Schulz}
\Name{{Schulz} T., {Ritz} R., {Bauer} A., {Halder} M., {Wagner} M., {Franz} C.,
  {Pfleiderer} C., {Everschor} K., {Garst} M. \and {Rosch} A.} \REVIEW{Nat.
  Phys.}{8}{2012}{301}.

\bibitem{Lange}
\Name{Lange H.} \REVIEW{physica status solidi (b)}{201}{1997}{3}.

\bibitem{Slack}
\Name{Slack G.~A.} \REVIEW{Materials Research Society, Pittsburgh,
  PA}{478}{1997}{47}.

\bibitem{Sales}
\Name{Sales B.~C., Jones E.~C., Chakoumakos B.~C., Fernandez-Baca J.~A., Harmon
  H.~E., Sharp J.~W. \and Volckmann E.~H.} \REVIEW{Phys. Rev.
  B}{50}{1994}{8207}.

\bibitem{V}
\Name{Jaccarino V., Wertheim G.~K., Wernick J.~H., Walker L.~R. \and Arajs S.}
  \REVIEW{Phys. Rev.}{160}{1967}{476}.

\bibitem{Wertheim}
\Name{Wertheim G.~K., Jaccarino V., Wernick J.~H., Seitchik J.~A., Williams
  H.~J. \and Sherwood R.~C.} \REVIEW{Phys. Lett.}{18}{1965}{89 }.

\bibitem{Watanabe}
\Name{Watanabe H., Yamamoto H. \and Ito K.-i.} \REVIEW{J. Phys. Soc.
  Jpn.}{18}{1963}{995}.

\bibitem{Moriya}
\Name{Takahashi Y., Tano M. \and Moriya T.} \REVIEW{J. Magn. Magn.
  Mater.}{31-34}{1983}{329 }.

\bibitem{Petrova}
\Name{Petrova A.~E., Krasnorussky V.~N., Shikov A.~A., Yuhasz W.~M., Lograsso
  T.~A., Lashley J.~C. \and Stishov S.~M.} \REVIEW{Phys. Rev.
  B}{82}{2010}{155124}.

\bibitem{Sakai}
\Name{{Sakai} A., {Yotsuhashi} S., {Adachi} H., {Ishii} F., {Onose} Y.,
  {Tomioka} Y., {Nagaosa} N. \and {Tokura} Y.} in proc. of \Book{26th
  International Conference on Thermoelectrics} 2007 p. 256.

\bibitem{Imada}
\Name{Imada M., Fujimori A. \and Tokura Y.} \REVIEW{Rev. Mod.
  Phys.}{70}{1998}{1039}.

\bibitem{Shu}
\Name{Ou-Yang T.~Y., Shu G.~J. \and Fuh H.~R.} \REVIEW{Europhys.
  Lett.}{120}{2017}{17002}.

\bibitem{Wolfe}
\Name{{Wolfe} R., {Wernick} J.~H. \and {Haszko} S.~E.} \REVIEW{Phys.
  Lett.}{19}{1965}{449}.

\bibitem{Chernikov}
\Name{Chernikov M.~A., Degiorgi L., Felder E., Paschen S., Bianchi A.~D., Ott
  H.~R., Sarrao J.~L., Fisk Z. \and Mandrus D.} \REVIEW{Phys. Rev.
  B}{56}{1997}{1366}.

\bibitem{Aarts}
\Name{F\"ath M., Aarts J., Menovsky A.~A., Nieuwenhuys G.~J. \and Mydosh J.~A.}
  \REVIEW{Phys. Rev. B}{58}{1998}{15483}.

\bibitem{Takagi}
\Name{Takagi S., Yasuoka H., Ogawa S. \and H.~Wernick J.} \REVIEW{Journal of
  the Physical Society of Japan}{50}{1981}{2539}.

\bibitem{Lisunov}
\Name{Lisunov K., Arushanov E., Kloc C., Broto J., Leotin J., Rokoto H.,
  Respaud M. \and Bucher E.} \REVIEW{Physica B: Condensed Matter}{229}{1996}{37
  }.

\bibitem{Mani2004}
\Name{Mani A.} \REVIEW{Solid State Communications}{132}{2004}{551 }.

\bibitem{Hearne}
\Name{Hearne G.~R., Musyimi P., Bhattacharjee S., Forthaus M.~K. \and
  Abd-Elmeguid M.~M.} \REVIEW{Phys. Rev. B}{100}{2019}{155118}.

\bibitem{Mani}
\Name{Mani A., Bharathi A. \and Hariharan Y.} \REVIEW{Phys. Rev.
  B}{63}{2001}{115103}.

\bibitem{Tomczak}
\Name{Tomczak J.~M., Haule K. \and Kotliar G.} \REVIEW{Proc. Natl Acad.
  Sci.}{109}{2012}{3243}.

\bibitem{Tomczak2018}
\Name{Tomczak J.~M.} \REVIEW{Journal of Physics: Condensed
  Matter}{30}{2018}{183001}.

\bibitem{Mattheiss}
\Name{Mattheiss L.~F. \and Hamann D.~R.} \REVIEW{Phys. Rev.
  B}{47}{1993}{13114}.

\bibitem{Aeppli}
\Name{Aeppli G. \and Fisk Z.} \REVIEW{Comment. Cond. Mat. Phys.}{16}{1992}{}.

\bibitem{Mazurenko}
\Name{Mazurenko V.~V., Shorikov A.~O., Lukoyanov A.~V., Kharlov K., Gorelov E.,
  Lichtenstein A.~I. \and Anisimov V.~I.} \REVIEW{Phys. Rev.
  B}{81}{2010}{125131}.

\bibitem{Dutta}
\Name{Dutta P. \and Pandey S.~K.} \REVIEW{J. Phys. Condens.
  Matter}{31}{2019}{145602}.

\bibitem{Doniach}
\Name{Fu C. \and Doniach S.} \REVIEW{Phys. Rev. B}{51}{1995}{17439}.

\bibitem{J}
\Name{Kune\'{s} J. \and Anisimov V.~I.} \REVIEW{Phys. Rev.
  B}{78}{2008}{033109}.

\bibitem{Klein}
\Name{Klein M., Zur D., Menzel D., Schoenes J., Doll K., R\"oder J. \and
  Reinert F.} \REVIEW{Phys. Rev. Lett.}{101}{2008}{046406}.

\bibitem{Beille}
\Name{{Beille} J., {Voiron} J. \and {Roth} M.} \REVIEW{Solid State
  Commun.}{47}{1983}{399}.

\bibitem{Bauer1998}
\Name{Bauer E., Galatanu A., Hauser R., Reichl C., Wiesinger G., Zaussinger G.,
  Galli M. \and Marabelli F.} \REVIEW{Journal of Magnetism and Magnetic
  Materials}{177-181}{1998}{1401 }.

\bibitem{Bauer}
\Name{Bauer E., Bocelli S., Hauser R., Marabelli F. \and Spolenak R.}
  \REVIEW{Physica B: Condensed Matter}{230}{1997}{794 }.

\bibitem{Sales2011}
\Name{Sales B.~C., Delaire O., McGuire M.~A. \and May A.~F.} \REVIEW{Phys. Rev.
  B}{83}{2011}{125209}.

\bibitem{Reichl}
\Name{Reichl C., Wiesinger G., Zaussinger G., Bauer E., Galli M. \and Marabelli
  F.} \REVIEW{Physica B: Condensed Matter}{259}{1999}{866 }.

\bibitem{Pelzer}
\Name{Pelzer R., Naber L., Galatanu A., Sassik H. \and Bauer E.}
  \REVIEW{Journal of Magnetism and Magnetic Materials}{226-230}{2001}{227 }.

\bibitem{Koyama}
\Name{Koyama K., Goto T., Kanomata T. \and Note R.} \REVIEW{Journal of the
  Physical Society of Japan}{68}{1999}{1693}.

\bibitem{Grechnev}
\Name{Grechnev G., Jarlborg T., Panfilov A., Peter M. \and Svechkarev I.}
  \REVIEW{Solid State Commun.}{91}{1994}{835 }.

\bibitem{Yee}
\Name{Haule K., Yee C.-H. \and Kim K.} \REVIEW{Phys. Rev. B}{81}{2010}{195107}.

\bibitem{kkr}
\REVIEW{}{}{}{} \url{http://kkr.issp.u-tokyo.ac.jp}.

\bibitem{Blaha}
\Name{Blaha P., Schwarz K., Madsen G. K.~H., Kvasnicka D. \and Luitz J.}
  \REVIEW{An augmented plane wave + local orbitals program for calculating
  crystal properties}{}{2001}{}.

\bibitem{Ark}
\Name{Bo\'{r}en B. \and Kemi A.} \REVIEW{Min. Geol.}{11A}{1933}{1}.

\bibitem{Jones}
\Name{Jones R.~O. \and Gunnarsson O.} \REVIEW{Rev. Mod. Phys.}{61}{1989}{689}.

\bibitem{Birch}
\Name{Birch F.} \REVIEW{Phys. Rev.}{71}{1947}{809}.

\bibitem{Anisimov}
\Name{Anisimov V.~I. \and Gunnarsson O.} \REVIEW{Phys. Rev. B}{43}{1991}{7570}.

\bibitem{Madsen}
\Name{{Madsen} G.~K.~H. \and {Nov{\'a}k} P.} \REVIEW{Europhys.
  Lett.}{69}{2005}{777}.

\bibitem{Paromita}
\Name{Dutta P. \and Pandey S.~K.} \REVIEW{Comput. Condens.
  Matter}{16}{2018}{e00325}.

\bibitem{Lal}
\Name{Lal S. \and Pandey S.~K.} \REVIEW{Phys. Lett. A}{381}{2017}{2117}.

\bibitem{Pandey}
\Name{Dutta P., Lal S. \and Pandey S.~K.} \REVIEW{Eur. Phys. J.
  B}{91}{2018}{183}.

\bibitem{S}
\Name{Dutta P., Lal S. \and Pandey S.~K.} \REVIEW{AIP Conf.
  Proc.}{1942}{2018}{090017}.

\bibitem{Birol}
\Name{Haule K. \and Birol T.} \REVIEW{Phys. Rev. Lett.}{115}{2015}{256402}.

\bibitem{K}
\Name{Haule K.} \REVIEW{Phys. Rev. B}{75}{2007}{155113}.

\bibitem{Haule}
\Name{Haule K.} \REVIEW{Phys. Rev. Lett.}{115}{2015}{196403}.

\bibitem{Paro}
\Name{Dutta P., Pandey S.~K. \and Lal S.} \REVIEW{Euro. Phys. J.
  B.}{91}{2018}{83}.

\bibitem{Kotliar}
\Name{Kotliar G., Savrasov S.~Y., Haule K., Oudovenko V.~S., Parcollet O. \and
  Marianetti C.~A.} \REVIEW{Rev. Mod. Phys.}{78}{2006}{865}.

\bibitem{dmft}
{\url{http://hauleweb.rutgers.edu/tutorials/whatis/whatis.html}}.

\bibitem{Jarrell}
\Name{Jarrell M. \and Gubernatis J.~E.} \REVIEW{Phys. Rep.}{269}{1996}{133}.

\bibitem{knittle}
\Name{Knittle E. \and Williams Q.} \REVIEW{Geophys. Res. Lett.}{22}{1995}{445}.

\bibitem{lin}
\Name{Lin J.~F., Campbell A., Heinz D.~L. \and Shen G.} \REVIEW{J. Geophys.
  Res.}{108(B1)}{2003a}{2045}.

\bibitem{wood}
\Name{Wood I.~G., David W. I.~F., Hull S. \and Price G.~D.} \REVIEW{J. Appl.
  Crystallogr.}{29}{1996}{215}.

\bibitem{guyot}
\Name{Guyot F., Zhang J., Martinez I., Matas J., Ricard Y. \and Javoy M.}
  \REVIEW{Eur. J. Mineral.}{9}{1997}{277}.

\bibitem{Zino}
\Name{Zinoveva G.~P., Andreeva L.~P. \and Geld P.~V.} \REVIEW{Phys. Status
  Solidi A}{23}{1974}{711}.

\bibitem{Spolenak}
\Name{Spolenak R., Bauer E., Müller H., Kirchmayr H., Blaha P., Mohn P. \and
  Schwarz K.} \REVIEW{Journal of Magnetism and Magnetic
  Materials}{157-158}{1996}{715 }.

\bibitem{Cooley}
\Name{Cooley J.~C., Aronson M.~C. \and Canfield P.~C.} \REVIEW{Phys. Rev.
  B}{55}{1997}{7533}.

\bibitem{Fang}
\Name{Fang Y., Ran S., Xie W., Wang S., Meng Y.~S. \and Maple M.~B.}
  \REVIEW{Proc. Natl. Acad. Sci.}{115}{2018}{8558}.

\bibitem{Yeo}
\Name{Yeo S., Nakatsuji S., Bianchi A.~D., Schlottmann P., Fisk Z., Balicas L.,
  Stampe P.~A. \and Kennedy R.~J.} \REVIEW{Phys. Rev. Lett.}{91}{2003}{046401}.

\bibitem{SP}
\Name{SupplementaryMaterial} \REVIEW{}{}{}{}.

\end{thebibliography}
\bibliographystyle{eplbib}

\end{document}



\title{Supplementary Material for: Understanding the temperature and pressure dependent electronic properties of FeSi: DFT+DMFT study}

\author{Paromita Dutta}
 \altaffiliation{dutta.paromita1@gmail.com}
 \affiliation{%
School of Basic Sciences, Indian Institute of Technology Mandi, Kamand, Himachal Pradesh-175075, India}%

\author{Sudhir K. Pandey}
\altaffiliation{sudhir@iitmandi.ac.in.com}
\affiliation{%
School of Engineering, Indian Institute of Technology Mandi, Kamand, Himachal Pradesh-175075, India
}%

\date{\today}



\maketitle

\begin{itemize}

\item[\textbf{S1.}] {\textbf{\textit{Temperature-dependent self-energy}}}

\end{itemize}
\small
\begin{figure}[tbh]
  \begin{center}
    \includegraphics[width=2.3in]{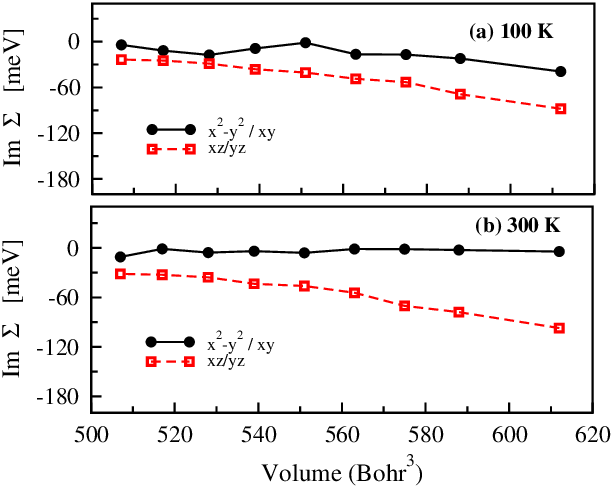}
  \end{center}

  \caption{\small{(color online) Imaginary parts of the self-energy (Im $\Sigma$) for orbital components x$^2$-y$^2$/xy and xz/yz of Fe 3\textit{d} states nearer to the Fermi level as a function of volume for (a) 100 K and (b) 300 K.}}
\end{figure} 

We have shown the plot of imaginary parts of self-energy (Im $\Sigma$) for orbital components x$^2$-y$^2$/xy and xz/yz of Fe 3\textit{d} states nearer to the Fermi level as a function of volume for 100 K and 300 K in Fig. 1(a) and 1(b), respectively. Im $\Sigma$ values for x$^2$-y$^2$/xy and xz/yz orbital components are calculated for $\omega \sim$ -0.11 eV and $\sim$ 0.23 eV, respectively. Here, it is important to note that only these orbitals are chosen on the basis of our last work \cite{Dutta}, where we found that these orbitals x$^2$-y$^2$/xy and xz/yz are contributing most, one in the VB and other in CB, respectively. The energies $\sim$ -0.11 eV and 0.23 eV correspond to the energy positions of the nearest peaks to the Fermi level corresponding to the orbitals one in VB and other in CB, respectively, as observed in our previous work \cite{Dutta}. Im $\Sigma$ provides the information regarding lifetime of quasiparticles states and they are related via uncertainty principle. More is the value of  Im $\Sigma$ more will be the lifetime broadening and more will be the quasiparticle-quasiparticle scatterings (for more details one can refer the ref. \cite{Imada}). 

In Fig. 1(a)-1(b) for 612 Bohr$^3$ volume, we found that with rise in temperature the value of Im $\Sigma$ for x$^2$-y$^2$/xy orbital component has been decreased (magnitude wise) from $\sim$ -40 meV to -5 meV for 100-300 K while for xz/yz it has come down to -97 meV from -87 meV. This indicates that the lifetime of quasiparticles corresponding to x$^2$-y$^2$/xy orbital is large as compared to the lifetime of quasiparticles corresponding to xz/yz orbital. Now, on observing for other volumes at 100 and 300 K, we found that with volume compression, the decreasing trend (magnitude wise) of Im $\Sigma$ is followed indicating enhancing the lifetime of quasiparticles for both the orbitals. Moreover, the quasiparticles lifetime are large at 300 K as compared to 100 K for both the orbitals at all the reduced volumes. Normally, it is said that larger value of lifetime of quasiparticles leads to lesser possibility of quasiparticles-quasiparticles scatterings. Thus, at 300 K the possibility of quasiparticles-quasiparticles scatterings appears to be less than at 100 K. 

\vspace{0.3in}

\begin{itemize}
\item[\textbf{S2.}]{\textbf{\textit{Temperature-dependent hybridization function}}}
\end{itemize}
\vspace{0.05 in}

\begin{figure}[tbh]

\subfloat[]{%
  \includegraphics[width=1.6in]{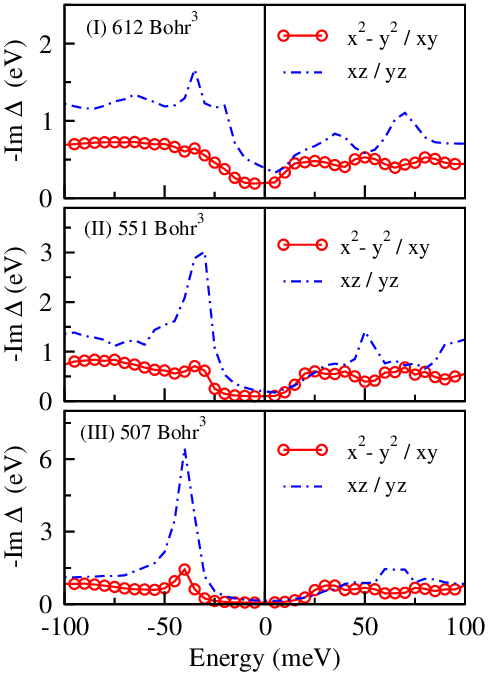}%
}\hfill
\subfloat[]{%
  \includegraphics[width=1.6in]{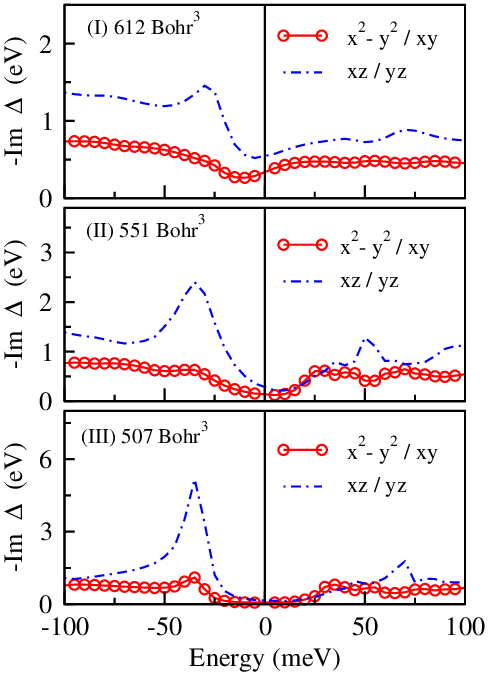}%
}
\caption{\small{(Colour online) The imaginary part of the impurity hybridization function for orbital components x$^2$-y$^2$/xy and xz/yz of Fe 3\textit{d} computed within DFT+DMFT \cite{Haule} for volumes \textit{viz.} (I) 612 Bohr$^3$, (II) 551 Bohr$^3$ and (III) 507 Bohr$^3$ at temperatures (a) 100 K and (b) 300 K, respectively.}}
\label{8}
\end{figure}

In this section we discuss the effect on the degree of localization of the correlated electrons under pressure with temperature inclusion on the basis of the impurity hybridization function for orbital components x$^2$-y$^2$/xy and xz/yz of Fe 3\textit{d} as computed within DFT+DMFT \cite{Haule}. Accordingly, the computed imaginary part of the impurity hybridization function (-Im $\Delta$) for volumes 612, 551 and 507 Bohr$^3$ are shown in Fig. 2 at temperatures \textit{viz.} (a) 100 K and (b) 300 K, respectively. Since no significant changes are found in the z$^2$ orbital component of Fe 3\textit{d}; we chose to not consider it in the figure. 

Considering Fig. 2(a), we found that at 100 K with volume compression in the valence band region (VBR) the energy peak of x$^2$-y$^2$/xy orbital component is becoming sharper while there is emergence of energy peak corresponding to xz/yz orbital component. For instance, at 612 Bohr$^3$ many small energy peaks corresponding to x$^2$-y$^2$/xy orbital are visible in VBR with one sharp peak positioned at $\sim$ -35 meV while a small hump is also seen corresponding to xz/yz orbital component. Here, the value of -Im $\Delta$ for the sharp peak is $\sim$ 1.6 eV. However, when we move to 551 Bohr$^3$ in VBR we found that only one peak corresponding to x$^2$-y$^2$/xy orbital positioned at $\sim$ -30 meV while hump corresponding to xz/yz orbital is now small peak at $\sim$ -33 meV. Here, the value of -Im $\Delta$ for the sharp peak has reached to $\sim$ 2.9 eV. Further on moving to 507 Bohr$^3$ in VBR, the energy peak corresponding to x$^2$-y$^2$/xy orbital has been sharpened and now positioned at $\sim$ -37 meV while the small peak corresponding to xz/yz orbital has now shifted to $\sim$ -40 meV. Finally, the value of -Im $\Delta$ for the sharp peak has reached to $\sim$ 6.3 eV. This implies at 100 K with volume compression the magnitude of -Im $\Delta$ is increasing. Generally, the information of the dynamics of correlated electrons hopping in and out of the bath is contained with the hybridization function. Large value (magnitude wise) of -Im $\Delta$ suggests the less localized electronic character of the system. On this basis, we can say that at 100 K the degree of localization of Fe 3\textit{d} electrons is found to be decreasing with volume reduction. Similar effect is noted at 300 K in Fig. 2(b), where sharpening of the energy peak corresponding to x$^2$-y$^2$/xy orbital has been occurred showing increased value of -Im $\Delta$ with volume reduction. Hence, it appears that volume compression is decreasing the degree of localization of Fe 3\textit{d} electrons in FeSi.  

On moving from Fig. 2(a)- 2(b), we observed few changes in the energy peak positions and the value of -Im $\Delta$ for both orbitals. In Fig. 2(b) at 612 Bohr$^3$ one broadened energy peak corresponding to x$^2$-y$^2$/xy orbital in VBR is observed positioned at $\sim$ -28 meV and its -Im $\Delta$ value has decreased to 1.4 eV. Then, at 551 Bohr$^3$ the broadened peak is now shifted to $\sim$ -33 meV and its value is again decreased to 2.4 eV as compared to 100 K data. Also, a small hump corresponding to xz/yz orbital can be seen here at $\sim$ -34 meV. Next moving to 507 Bohr$^3$, we noticed that the broadened peak has now changed to sharp peak while its -Im $\Delta$ value is still smaller than its value as found at 100 K by 3.9 eV. It is found that -Im $\Delta_{300K} <$ -Im $\Delta_{100K}$ (magnitude wise) for all the reduced volumes. Thus, at 300 K the Fe 3\textit{d} electrons are more localized than 100 K. Hence, we can conclude that volume compression will decrease the degree of localization of Fe 3\textit{d} electrons while temperature rise seems to localize them.